\documentclass[journal,twoside,web]{ieeecolor}

\usepackage{silence}
\usepackage[font=scriptsize]{caption}
\usepackage{array}
\usepackage{dirtytalk}
\usepackage{algorithm} 
\usepackage{algpseudocode} 
\usepackage{soul}

\usepackage{generic}
\usepackage{amsmath,amssymb,amsfonts}

\usepackage{graphicx}
\usepackage{textcomp}
\usepackage{booktabs}
\usepackage{multirow}
\usepackage{hyperref}

\usepackage{enumerate}
\usepackage{pifont}
\usepackage{fancyhdr}
\fancypagestyle{FirstPage}{

\lhead{This work is under review. Copyright may be transferred without notice, after which this version may no longer be accessible.}
\lfoot{Email: ussman@snu.ac.kr}

}

\def\BibTeX{{\rm B\kern-.05em{\sc i\kern-.025em b}\kern-.08em

    T\kern-.1667em\lower.7ex\hbox{E}\kern-.125emX}}

\begin{document}


\title{MESAHA-Net: Multi-Encoders based Self-Adaptive Hard Attention Network with Maximum Intensity Projections for Lung Nodule Segmentation in CT Scan}

\author{Muhammad Usman, Azka Rehman, Abd~Ur~Rehman, Abdullah Shahid, Tariq Mahmood Khan, Imran Razzak, Minyoung Chung, and Yeong‑Gil Shin
\thanks{Manuscript received March 31, 2023; (Corresponding author: Muhammad Usman.)}
\thanks{Muhammad Usman, Minyoung Chung, and Yeong-Gil Shin are with the Department of Computer Science and Engineering, Seoul National University, 1 Gwanak-ro, Gwanak-gu, Seoul 08826, Republic of Korea, (e-mail: ussman@snu.ac.kr; yshin@snu.ac.kr).}
\thanks{Muhammad Usman, Azka Rehman, and Abdullah Shahid is with Center for Artificial Intelligence in Medicine and Imaging, HealthHub Co. Ltd., Seoul 06524, Republic of Korea, (e-mail: usman@healthhub.kr; azka@healthhub.kr; abdullah@healthhub.kr; ).}
\thanks{Abd~Ur~Rehman is with Department of Electrical and Computer Engineering, The University of Alabama, AL 35401, United States.}
\thanks{Tariq Mahmood Khan and Dr. Imran Razzak is with Department of Electrical Engineering, COMSATS University Islamabad, Islamabad, Pakistan, (e-mail: tariq\_mehmood@comsats.edu.pk)}}

\maketitle

\begin{abstract}
Accurate lung nodule segmentation is crucial for early-stage lung cancer diagnosis, as it can substantially enhance patient survival rates. Computed tomography (CT) images are widely employed for early diagnosis in lung nodule analysis. However, the heterogeneity of lung nodules, size diversity, and the complexity of the surrounding environment pose challenges for developing robust nodule segmentation methods. In this study, we propose an efficient end-to-end framework, the multi-encoder-based self-adaptive hard attention network (MESAHA-Net), for precise lung nodule segmentation in CT scans. MESAHA-Net comprises three encoding paths, an attention block, and a decoder block, facilitating the integration of three types of inputs: CT slice patches, forward and backward maximum intensity projection (MIP) images, and region of interest (ROI) masks encompassing the nodule. By employing a novel adaptive hard attention mechanism, MESAHA-Net iteratively performs slice-by-slice 2D segmentation of lung nodules, focusing on the nodule region in each slice to generate 3D volumetric segmentation of lung nodules. 
The proposed framework has been comprehensively evaluated on the LIDC-IDRI dataset, the largest publicly available dataset for lung nodule segmentation. The results demonstrate that our approach is highly robust for various lung nodule types, outperforming previous state-of-the-art techniques in terms of segmentation accuracy and computational complexity, rendering it suitable for real-time clinical implementation.

\end{abstract}


\section*{Introduction}
\label{intro}
\thispagestyle{FirstPage}

Lung cancer is the leading cause of death worldwide, accounting for approximately 1.8 million deaths annually \cite{bray2018global}. Early detection and diagnosis of lung cancer are crucial for improving patient outcomes, including prognosis and survival rates \cite{henschke2006survival}. Computed tomography (CT) is the preferred imaging modality for monitoring treatment response and analyzing pulmonary nodules \cite{aerts2014decoding}. While most lung nodules are benign, approximately 20-30\% of lung nodules detected on CT scans can be malignant and indicative of early-stage lung cancer \cite{gould2007evaluating}. The accurate segmentation of lung nodules is crucial in assisting the early diagnosis of lung cancer. However, the task of segmenting lung nodules is challenging due to the heterogeneity of nodule types, size variations, and the complexity of the surrounding lung anatomy \cite{armato2011lung}. CT scans are extensively utilized for the diagnosis of lung nodule has emerged as it has proven the most effective imaging modality for early diagnosis and analysis \cite{henschke2006survival}.

\begin{figure}[ht]
\centering
\centerline{\includegraphics[width=0.45\textwidth]{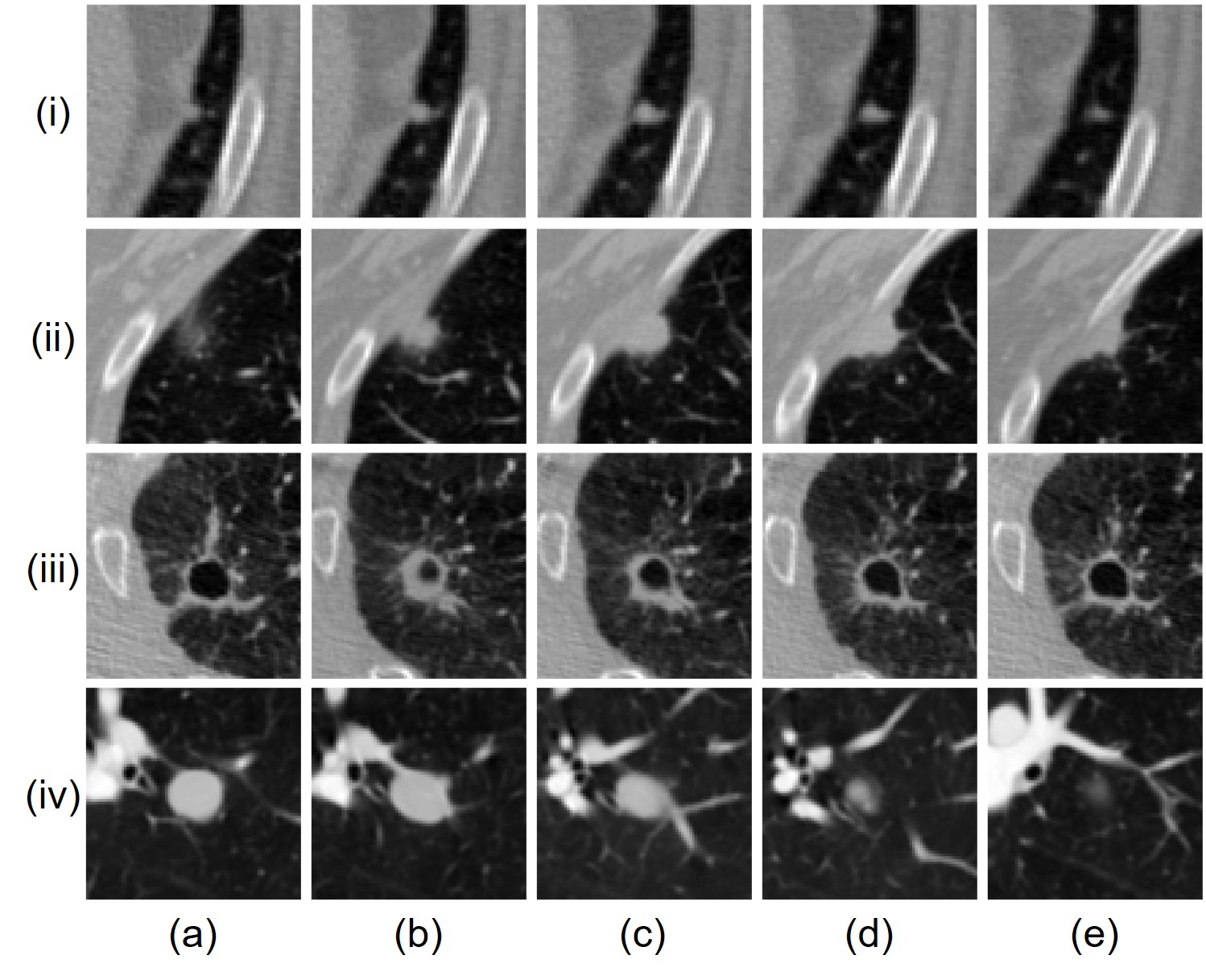}}
\caption{Multiple visual appearances of the pulmonary nodule are shown. The variations within a nodule can be seen from column (a) to (e), and the inter-nodule difference is presented from rows (i) to (iv).}
\label{nodule_type}
\end{figure}

Despite the advances in CT imaging and its role in early lung cancer detection, manual segmentation of lung nodules remains a time-consuming and labor-intensive process, with several disadvantages and limitations. Manual segmentation is subject to intra- and inter-observer variability, leading to inconsistent and unreliable results \cite{mcnitt-gray2009standardizing}. Additionally, the large number of CT images generated in lung cancer screening programs presents a significant workload for radiologists, increasing the likelihood of human error and missed or misinterpreted nodules \cite{giger2013computer}. Furthermore, the subtle and diverse nature of lung nodules can also pose challenges for manual segmentation, particularly in the presence of complex anatomical structures, such as blood vessels and bronchi \cite{armato2011lung} as shown in Fig. \ref{nodule_type}. These limitations of manual segmentation highlight the need for the development of computer-aided diagnosis (CAD) systems for the segmentation and analysis of lung nodules. 


A multitude of efforts have been dedicated to the development of computer-aided diagnosis (CAD) systems to aid radiologists in identifying and analyzing lung nodules from CT scans \cite{setio2016pulmonary}. These CAD systems aim to enhance the diagnostic process by automating the identification and segmentation of lung nodules, thereby reducing the workload of radiologists and minimizing the potential for human error \cite{mcwilliams2013probability}.
Various techniques have been employed in the development of computer-aided diagnosis (CAD) systems for lung nodule segmentation, ranging from traditional image processing approaches \cite{lassen2015robust,diciotti2011automated,song2015lung} and machine learning algorithms \cite{lu2014computer, gao2016dropout}, to deep learning (DL)-based methods \cite{hamidian20173d, fu2017automatic}. Among these techniques, DL-based methods have demonstrated superior performance compared to conventional approaches, exhibiting enhanced segmentation accuracy and robustness against diverse lung nodule characteristics, such as type, texture, and size \cite{warrier2023review}. 



Deep learning (DL)-based studies for lung nodule segmentation can be broadly categorized into three groups: voxel-level classification (VLC) \cite{wang2017central,cao2020dual,liu2019cascaded}, 3D segmentation \cite{zhou2022cascaded,zhang2022multi,tyagi2022cse}, and 2D patch-wise segmentation \cite{usman2020volumetric,wu2021coarse,chen2021mtgan}. The VLC methodology involves classifying each voxel within the volume of interest (VOI) as either nodular or non-nodular. However, this approach fails to exploit global semantic information, leading to an increased number of false positives stemming from inter-class similarities at nodule boundaries. Moreover, the substantial computational resources demanded by VLC methods render them ill-suited for real-time clinical applications.
Several studies have approached lung nodule segmentation as a 3D segmentation problem, wherein the entire volume of interest (VOI) is input to the network for generating a 3D nodule segmentation \cite{zhou2022cascaded,zhang2022multi,tyagi2022cse}. However, these methodologies are plagued by suboptimal segmentation performance due to rescaling-induced error (RIE), which arises from significant rescaling during pre- and post-processing stages
Given the variability in nodule sizes, determining the optimal input dimension for 3D networks proves challenging, consequently leading to a substantial decline in segmentation outcomes. Furthermore, 3D models necessitate considerable computational resources for both training and deployment in clinical environments.



To address the limitations associated with VLC and 3D segmentation, 2D patch-wise segmentation strategies are proposed in \cite{usman2020volumetric,wu2021coarse,chen2021mtgan}. These techniques perform 3D segmentation of nodules using 2D regions of interest (ROIs) as input, encompassing the nodule in each slice. To estimate ROIs in adjacent slices, two types of strategies have been employed: constant/fixed ROIs \cite{liu2019cascaded} and dynamic ROIs methods \cite{usman2020volumetric}. In constant or fixed ROI-based methods, the same ROI is applied to all surrounding slices, introducing redundant regions in the ROIs and posing challenges for the network during nodule segmentation. In contrast, dynamic ROI-based approaches determine ROIs for adjacent slices dynamically, based on the segmentation of the nodule in the current slice, effectively eliminating redundant regions and enabling the network to concentrate on nodular areas.

Although these techniques have significantly improved performance through multi-view analysis, they remain susceptible to RIE, limiting the robustness of such approaches. To tackle these challenges, in this study, we introduce a novel multi-encoder-based self-adaptive hard attention network (MESAHA-Net) for volumetric segmentation of pulmonary nodules. The proposed framework utilizes only the 2D ROI as input to yield 3D segmentation of a given lung nodule. MESAHA-Net comprises three encoding paths, an attention block, and a single decoding block. The proposed architecture incorporates three inputs: the cropped slice, as well as forward and backward Maximum Intensity Projection (MIP) images with a thickness of 3 mm, which are employed to segment the nodule and estimate the ROI for forward and backward slices, respectively. Moreover, ROI information is supplied as hard attention to the attention block, allowing the network to concentrate solely on the ROI within the given slice patch. This enables the network to tackle the challenges posed by nodule size diversity. The network also leverages spatial and contextual information to segment the nodule in the provided slice and estimate ROIs for adjacent slices, which are subsequently used as input for nodule segmentation in later iterations.

\section*{Related Work}
\label{RW}
Various efforts have been devoted to the development of highly accurate and robust automatic nodule segmentation systems to aid radiologists \cite{gu2021survey}. These approaches are typically designed to identify the nodular voxels in the given volume of interest (VOI) containing the nodule. The most commonly employed traditional techniques in this context include morphological method \cite{song2015lung}, region growing method \cite{diciotti2011automated}, and energy optimization method \cite{lassen2015robust}. However, several studies have compared traditional algorithms with advanced deep learning-based techniques for lung nodule segmentation, demonstrating the superior performance of the latter in terms of accuracy and robustness \cite{cao2017improving,wang2017central}. 


Deep learning-based methodologies have addressed the lung nodule segmentation challenge through various approaches, including voxel-level classification (VLC), 2D patch-wise segmentation, and 3D segmentation techniques. For example, Wang et al. \cite{wang2017central} introduced Central Focused Convolutional Neural Networks (CF-CNN) with central pooling, employing both 2D and 3D inputs surrounding a specific voxel for nodule and non-nodule classification. Likewise, Cao et al. \cite{cao2020dual} proposed a central intensity pooling layer (CIP) to extract intensity features from the center voxel of a given volume of interest (VOI), subsequently utilizing a CNN to obtain convolutional features for voxel classification. Similarly, Liu et al. \cite{liu2019cascaded} suggested a dual-pathway architecture, comprising ResBlocks, to integrate 2D and 3D inputs for voxel classification. However, voxel-level classification techniques often fail to optimally exploit global semantic information, resulting in a high false positive rate. Additionally, the limited generalization capabilities of these models, coupled with the substantial inference time required given VOI, render them unsuitable for clinical applications.


Several studies have proposed 3D segmentation-based approaches to address issues related to VLC. In such approaches, the entire VOI is inputted to generate a 3D segmentation of the given lung nodule. However, this approach faces challenges due to the varied size of lung nodules, which make it difficult to select the optimal VOI dimensions. To overcome this problem, Zhou et al. \cite{zhou2022cascaded} developed a lung nodule segmentation system that utilized a 3D UNet architecture. They added the Atrous Spatial Pyramid Pooling (ASPP) module and convolutional block attention module (CBAM) to fuse high-level features in the multi-scale and supplement global features, respectively. Similarly, Zhang et al. \cite{zhang2022multi} trained four 3D UNet networks of different input dimensions to overcome the scale variant issue of lung nodules. They ensemble the output of all four networks using multi-scale dense CRF to produce the 3D volumetric lung nodule mask. Likewise, Tyagi et al. \cite{tyagi2022cse} employed a 3D conditional generative adversarial network, consisting of a 3D UNet and a classification network, with a spatial squeeze and channel excitation modules, that acted as a generator and discriminator, respectively. These approaches achieved significantly better performance, but they incur significant computational complexities, making them impractical for real-time clinical utilization.

To overcome the issues pertaining to the computational complexities, 2D patch-wise approaches were followed by a number of studies. For instance, Usman et al. \cite{usman2020volumetric} proposed a novel method that utilized 2D ROI input to generate 3D lung nodule segmentation. They employed a residual U-Net architecture for nodule segmentation in multi-view analysis (MVA) and introduced an innovative adaptive ROI (A-ROI) algorithm to determine the ROI for each successive slice based on the current slice's nodule segmentation mask. Wu et al. \cite{wu2021coarse} proposed a patch-wise segmentation scheme comprising two U-Nets to accommodate both 2D and 3D patch inputs. The outputs of both networks were combined to generate nodule segmentation for the given slice. Chen et al. \cite{chen2021mtgan} introduced a novel Mask and Texture-driven Generative Adversarial Network (MTGAN) with a joint multi-scale L1 loss for lung nodule segmentation. The MTGAN employed a U-Net as the generator and a five-layer fully convolutional network (FCN) as the discriminator. Similarly, Usman et al. \cite{usman2022deha} proposed a dual-encoder-based architecture that eliminated the need for rescaling during pre- and post-processing stages. They introduced hard attention to counteract issues caused by rescaling and utilized the A-ROI algorithm for examining the nodule in its surroundings. These techniques achieved significant improvements and robustness in lung nodule segmentation performance. However, due to the utilization of complex network architectures, the implementation of MVA, and the use of handicraft algorithms, the overall complexity of these solution was substantially increased.

The primary differences between the proposed MESAHA-Net and previous approaches are three-fold: 1) we introduced a lightweight multi-encoder-based architecture capable of leveraging the ROI mask as hard attention to produce highly accurate lung nodule segmentation efficiently; 2) we incorporated forward and backward MIP images to effectively estimate the ROI for adjacent slices without employing any handcrafted algorithm; 3) we eliminated the need for rescaling at pre- or post-processing stages by utilizing a novel adaptive hard attention mechanism which eliminates the rescaling-induced error. 


\section*{Proposed Method}
\label{pro}
The proposed multi-encoder based self-adaptive hard attention network (MESAHA-Net) is 
\begin{figure*}[ht]
\centering
\centerline{\includegraphics[width=\textwidth]{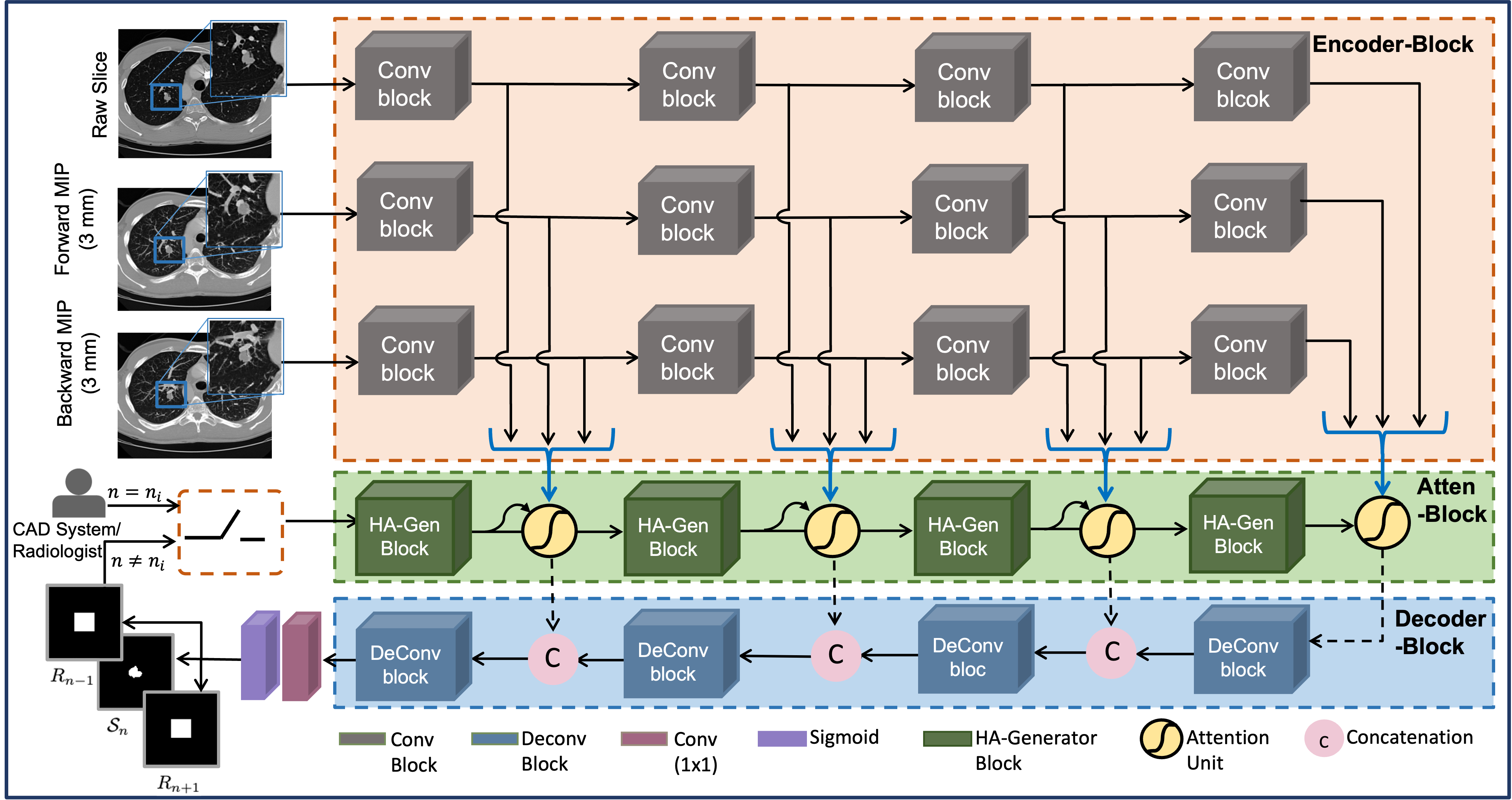}}
\caption{The overall architecture of proposed multi-encoder self-adaptive hard attention network (MESAHA-Net) for Lung Nodule segmentation is demonstrated.}
\label{propose_net}
\end{figure*}
\begin{figure}[hb]
\centering
\centerline{\includegraphics[width=0.44\textwidth]{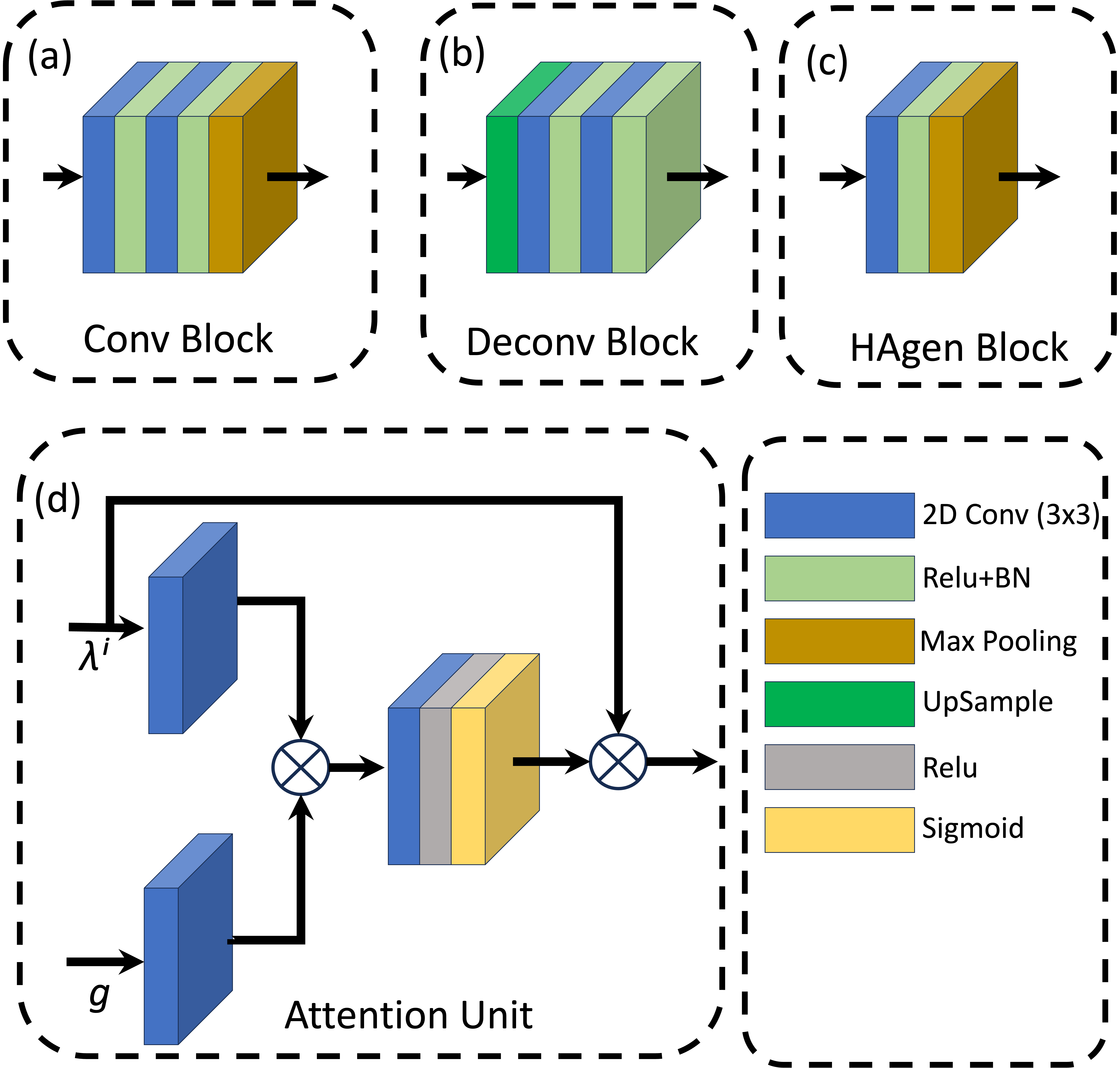}}
\caption{Illustration of blocks utilized in MESAHA-Net: (a) Convolutional Block (Conv-Block), (b) Deconvolutional Block (Deconv-Block), (c) Attention Generator Block (AGen Block), and (d) Attention Unit (AU).}
\label{network_units}
\end{figure}
equipped with three encoding path blocks, a hard-attention block, and a decoder block. The three encoding paths allow the network to integrate slice patch, forward, and backward MIP images, providing spatial and 3D contextual information to enhance nodule segmentation performance and ROI estimation for adjacent slices. The attention block facilitates the incorporation of the ROI mask as hard attention, directing the network's focus toward the nodular region. Lastly, the decoder block processes the refined information to generate the nodule segmentation mask and ROI masks for adjacent slices. Detailed descriptions of each block are provided in the following sections.

\subsection{Encoder block}
The encoder block comprises three encoding branches, with each accepting a single 2D image input. Specifically, the raw slice patch, and 3 mm thick forward and backward MIP images are fed into each encoder. The raw slice patch provides low-level information and 2D context, crucial for accurately segmenting the nodule boundary within the given ROI. Forward and backward MIP images compensate for the 3D spatial aspect by offering insights into the nodule from both sides of the given slice.

All three encoders share an identical architecture. Each encoder is composed of four Convolutional Blocks (Conv-blocks), which consist of two sets of 2D convolutions as shown in Fig. \ref{network_units} (a). Each set includes a convolution layer followed by ReLU activation and batch normalization layers, with max pooling applied afterward. Conv-blocks downsample the input while extracting coarse features related to the nodule. These features are fed to the attention block using skip connections, and the attention block forwards the refined features to the decoder block for further processing.

\subsection{Self-Adaptive Hard Attention Mechanism}
The proposed MESAHA-Net employs a hard attention mechanism, utilizing a ROI mask for initial segmentation. This network generates analogous ROI masks for adjacent slices, which are subsequently fed back into the system for iterative nodule segmentation in successive slices. A detailed description of the self-hard attention mechanism's algorithmic steps can be found in Algorithm \ref{algo_1}.

\begin{algorithm} 
        
	\caption{: Self-Adoptive Hard Attention Mechanism}
	\begin{algorithmic}[1]
            \State $n = n_i$, $R_n = R_{n_i}$
            \While {$\sum{R_n} > 0$}
            \State $\mathcal{I}_{\text{n}} = iCrop(I_{\text{n}}, R_n)$
            \State $\mathcal{M}^{\text{-}} = max(\mathcal{I}_{\text{n-2}}, \mathcal{I}_{\text{n-1}}, \mathcal{I}_{\text{n}})$
            \State $\mathcal{M}^{\text{+}}= max(\mathcal{I}_{\text{n}}, \mathcal{I}_{\text{n+1}}, \mathcal{I}_{\text{n+2}})$
            \State $R_{n-1}, \mathcal{S}_{n}, R_{n+1} = \mathcal{\psi}(\mathcal{I}_n, \mathcal{M}^{\text{-}}, \mathcal{M}^{\text{+}}, R_n)$
            \State $n \leftarrow n \pm 1$
            \State $R_n \leftarrow R_{n\pm1}$
            \EndWhile
	\end{algorithmic}
	\label{algo_1}
\end{algorithm}

In Algorithm \ref{algo_1}, $n_i$ and $R_{n_i}$ represent the inputted slice number and its corresponding ROI mask, respectively. The proposed algorithm initially extracts a $96 \times 96$ patch surrounding the given ROI, denoted by $\mathcal{I}_n$. Subsequently, this cropped patch is employed to generate forward and backward MIP images, represented as $\mathcal{M}^{\text{-}}$ and $\mathcal{M}^{\text{+}}$, respectively. These MIP images are produced with a 3 mm slab thickness by utilizing three adjacent slices.

The MESAHA-Net, denoted by $\mathcal{\psi}$, accepts four inputs: the cropped patch of the slice, forward and backward MIP images, and the ROI mask. It predicts the lung nodule segmentation mask, $S_n$, for the $n^{th}$ slice, along with ROI masks for both adjacent slices, i.e., $R_{n-1}$ and $R_{n+1}$. To examine the presence of nodules in neighboring slices, the algorithm employs the predicted ROI masks as input for the next iteration. Segmentation is executed in both directions to obtain the volumetric segmentation of the given nodule.


\subsection{Attention Block}
In the encoder block, convolutional layers systematically extract meaningful features ($\lambda^i$) from high-dimensional inputs, such as raw slice patch, forward and backward 3 mm MIP images. This sequential process enables the model to capture a substantial receptive field and the semantic contextual information vital for segmenting lung nodules in the given slice. However, features derived from a large receptive field contain an abundance of redundant information that does not correspond to the actual nodule's region of interest.
To address this issue, an attention block is incorporated into MESAHA-Net to refine the coarse features extracted from each encoder branch after every conv-block. This attention block receives coarse features and filters them to suppress redundant features by assigning lower weights. These refined features are then forwarded to the decoder block for further processing.


The architecture of attention block has been demonstrated in Figure \ref{propose_net}. The block consists of hard attention generator (HA-Gen) units and attention gating (AG) units. The role of HA-Gen units is to exploit the inputted ROI mask to generate the gating vector $g_j$ for AG units. The AG unit consists of 2D convolution layer followed by batch normalization, relu and max-pooling layers as shown in Fig. \ref{network_units} (c). Whereas the overall architecture of AG unit is inspired from \cite{oktay2018attention}, however, in contrast to \cite{oktay2018attention}, we employed multiplicative attention \cite{mnih2014recurrent} owing to its computational efficiency over additive attention \cite{bahdanau2014neural}. The output of AG ($\hat{\lambda}^i_{j,c}$) units is the element-wise multiplication of input features-maps ($\lambda^i_{j}$) extracted from $ith$ convolution block and hard-attention coefficients $\mathcal{H}^i_{j}$ and can be defined as follows: 

\begin{equation}
    \hat{\lambda}^i_{j,c}= \lambda^i_{j,c} . \mathcal{H}^i_{j} 
\end{equation}

Here hard attention coefficients, $\mathcal{H}^i_j \in [0,1] $, recognize the salient image region which is essentially in our case is the ROI containing the lung nodule. Where $j$ and $c$ represent the spatial and channel dimensions, respectively. $\mathcal{H}^i_j $ can be defined as follows: 

\begin{equation}
    \mathcal{H}^i_j = \sigma (\rho^i_{HA}(\lambda^i_{j,c},g_j;\theta_{att}))
\end{equation}

\begin{equation}
    \rho^i_{HA} = \eta^{\tau} (\xi ( W^{\tau}_{\lambda} \lambda^i_j  .  W_g^{\tau} g_j + \beta_g ) ) + \beta_{\eta} 
\end{equation}

\begin{equation}
    \sigma(\lambda^i_{j,c}) = \frac{1}{1 + e^{-\lambda^i_{j,c}}} 
\end{equation}

In other words, attention unit is characterised by a set of parameters $\theta_{att}$ containing; linear transformations $W_\lambda \in \mathbb{R}^{F_l \times F_{\text {int }}}, W_g \in \mathbb{R}^{F_g \times F_{\text {int }}}, \eta \in \mathbb{R}^{F_{\text {int }}\times 1}$ and bias terms  $\beta_g \in \mathbb{R}, \beta_{\eta} \in \mathbb{R}^{F_{\text {int }}}$.  The linear transformations are computed using channel-wise $1\times 1\times 1$ convolutions for the input tensors. This can be regarded as the vector concatenation-based attention, where the concatenated features $\lambda^i$ and $g$ are linearly mapped to a $\mathbb{R}^{F_{\text {int }}}$ dimensional intermediate space.

\subsection{Decoder block}
The decoder block is designed to process refined features from four levels of attention blocks, as illustrated in Fig. \ref{propose_net}. This block primarily upsamples inputs and sequentially structures feature maps for nodule segmentation and ROI masks. Its architecture comprises Deconv blocks, as depicted in Fig. \ref{network_units}(b).Each Deconv block first upsamples the input, followed by two sets of 2D convolutions with ReLU activation. Afterward, a batch normalization layer is applied, followed by a second CNN layer with ReLU activation. The feature map at this stage is combined with the input from the ConvBlock, and an upsampling layer is applied subsequently. Every ConvBlock from the first three encoders is concatenated and passed to the attention units, with a feature map from the attention-generating units serving as the gating signal. These attention units' outputs are concatenated with the DeConv block outputs of the decoder.

\subsection{Loss Function}

The total loss we used to train the model is combination of two losses, i.e., attention and segmentation loss.

\begin{equation}
\label{totalloass}
\mathcal{L}_{Total} = \alpha*\mathcal{L}_{\text{Seg}}+ (1-\alpha) \mathcal{L}_{Att},
\end{equation}

Here $\alpha$ is dynamic hyper parameter which is used to control the contribution of each loss and can defined as follows:

\begin{equation}
\begin{aligned}
\alpha & =\left\{\begin{array}{l}
1- \tau, \sum R_{n\pm1} = 0 \\
\tau, \text { otherwise }
\end{array}\right.
\end{aligned}
\end{equation}

Since the proposed MESAHA-Net is 2D segmentation model and performs slice-by-slice nodule segmentation in axial view, it is crucial for network to accurately determine the ends of nodule along z-axis. Therefore, we use lower values of $\alpha$ for end slices from both slides and higher for the intermediate slices. End slices are defined as those slices for which any of hard attention, ROI mask, becomes completely blank as there is no further nodule present in the next coming slice. The purpose is to enforce the network to learn the absence of nodule in the upcoming slice to stop the further investigation in corresponding direction. We set $\tau$ to 0.3 for training the proposed MESAHA-Net. 

Attention loss is computed by calculating the dice loss between predicted ROI masks ($R_{n\pm1}^p$) by MESAHA-Net and ideal ROI masks ($R_{n\pm1}^g$) for $n\pm1th$ slice. Ideal ROI mask is generated by using the reference nodule masks while keeping the margin 20\% from all fourth sides. Attention loss can defined as: 

\begin{equation}
\label{Atttotalloass}
\mathcal{L}_{\text{Att}} =  \left[ 1 - \frac{R_{n+1}^p\cap R_{n+1}^g}{R_{n+1}^p + R_{n+1}^g} + \frac{R_{n-1}^p\cap R_{n-1}^g}{R_{n-1}^p + RoI_{n-1}}\right]
\end{equation} 



The goal of nodule segmentation is to precisely detect the boundary/edges of nodule and to achieve this goal we employed the boundary loss \cite{Kervadec_Boundary_2021} which is computed by integral framework that approximates the distance by computing the integrals over the interface between mismatch regions of the two boundaries. To calculate the distance Dist$(\partial G, \partial S)$ between two boundaries of ground truth mask ($\partial G$) and predicted mask ($\partial S$), they used the integral framework to formulate the boundary loss by formulating as follows :

\begin{align}
\operatorname{Dist}(\partial G, \partial S) & =\int_{\partial G}\left\|m_{\partial S}(p)-p\right\|^2 d p \nonumber \\
& = 2 \int_{\Delta S} D_G(p) d p \nonumber \\
& \left.=2\left(\int_{\Omega} \phi_G(p) s(p) d p-\int_{\Omega} \phi_G(p) g(p) d p\right)\right) \label{eq:distance}
\end{align}

Let $p$ be a point on the boundary $\partial G$ and $m_{\partial S}(p)$ be the corresponding point on the boundary $\partial S$. The distance map of point $q$ concerning the boundary $\partial G$ is represented by $D_G(p)$. In this case, $\Omega$ denotes the spatial domain, and binary indicator functions are used. The binary indicator function of region $S$, denoted by $s: \Omega \rightarrow{{0,1}}$, is defined as $s(q)=$ 1 if $q \in S$ belongs to the target and 0 otherwise. The level set representation of boundary $\partial G$ is represented by $\phi_G: \Omega \rightarrow \mathbb{R}$, such that $\phi_G(q)=-D_G(q)$ if $q \in G$ and $\phi_G(q)=D_G(q)$ otherwise. By considering $S=S_\theta$ and replacing the binary variables $s(q)$ in Eq. (4) with the softmax probability outputs of the network $s_\theta(q)$, we derive the following boundary loss. This loss, up to a constant independent of $\theta$, approximates the boundary distance $\operatorname{Dist}\left(\partial G, \partial S_\theta\right)$:

$$
\mathcal{L}_{Seg}(\theta)=\int_{\Omega} \phi_G(q) s_\theta(q) d q
$$



\begin{figure}[t]
\centering
\centerline{\includegraphics[width=.35\textwidth]{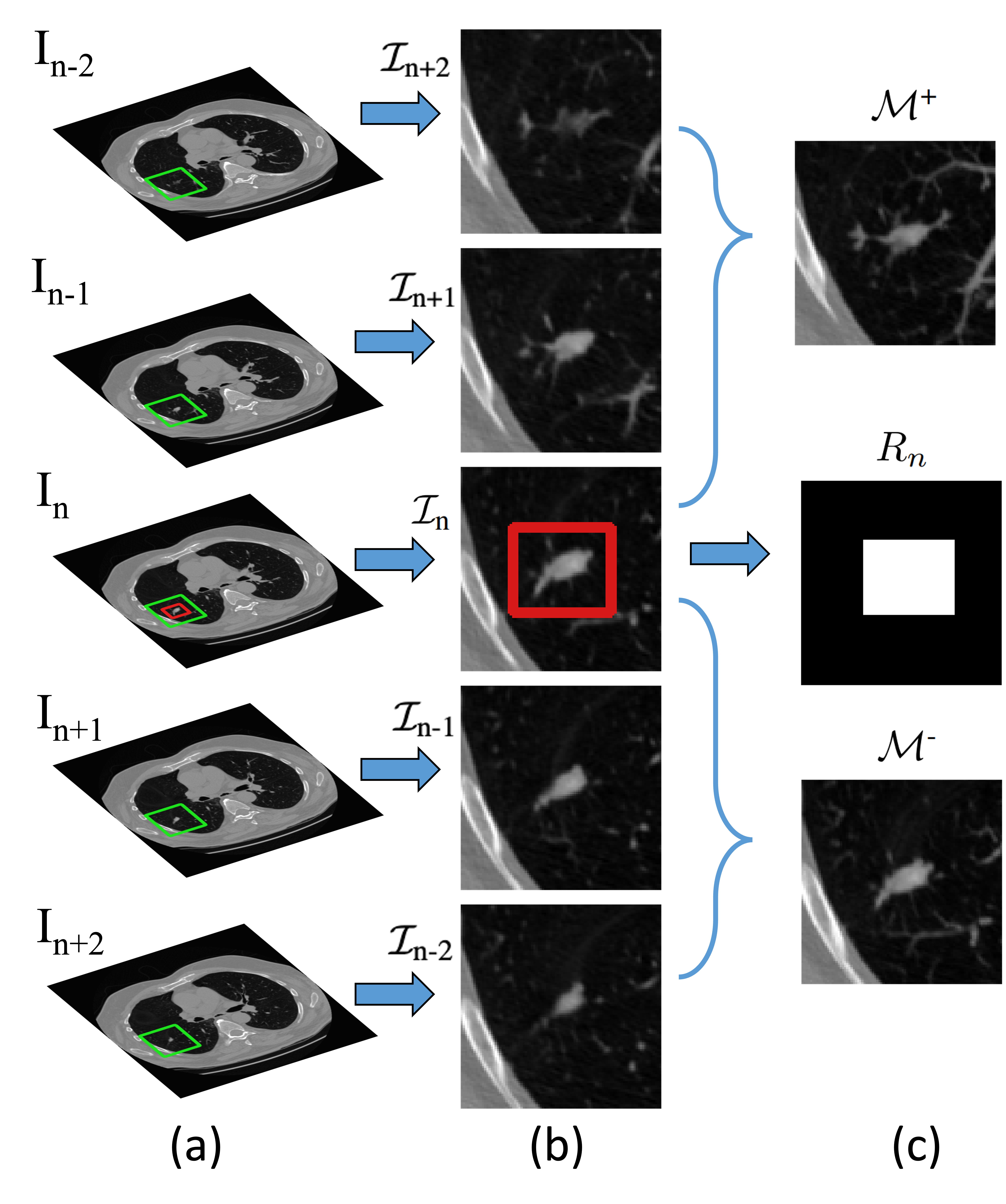}}
\caption{The preprocessing stage is illustrated as follows: (a) Patch position selection on the slices using the provided ROI, with the selected patch and ROI highlighted in green and red, respectively. (b) Cropped patches. (c) Forward and backward MIP images along with the ROI mask.}
\label{pre_processing}
\end{figure}
\section{Experimental Setup}

\subsection{Dataset and Preprocessing}
We evaluated the performance of our proposed MESAHA-Net using the Lung Image Database Consortium and Image Database Resource Initiative (LIDC/IDRI) dataset \cite{armato2011lung,setio2016pulmonary}. This dataset contains 1,018 thoracic CT scans with marked-up annotated lesions from 1,010 patients, sourced from seven universities and eight medical imaging companies globally. It encompasses 2,610 lung nodules with diameters ranging from 2.03 mm to 38.12 mm, slice thicknesses varying between 0.45 mm and 5.0 mm, and axial plane resolutions (pixel spacing) from 0.46 mm $\times$ 0.46 mm to 0.98 mm $\times$ 0.98 mm. Four specialized radiologists provided the nodule annotations.

For our study, we selected nodules annotated by all four experts, totaling 893 nodules. We employed a 50\% consensus criterion to establish ground-truth boundaries, addressing the inter-observer variability among the radiologists \cite{kubota2011segmentation}. These expert radiologists evaluated all 893 nodules and provided annotations for nine characteristics: Subtlety, Internal Structure, Calcification, Sphericity, Margin, Lobulation, Speculation, Texture, and Malignancy \cite{armato2011lung}. Following previous studies \cite{wang2017central, usman2020volumetric}, we partitioned the 893 nodules into training, validation, and testing sets, containing 356 (40\%), 45 (5\%), and 492 (55\%) nodules, respectively. Table \ref{LIDC_data_dist} demonstrates that the distribution of clinical characteristics among the three subsets is statistically comparable. We trained the proposed MESAHA-Net exclusively on the training set by utilizing the validation set to determine the number of training epochs. Lastly, we employed the test set to assess the performance of our framework.

\begin{table}{!ht}
\begin{center}
\caption{Characteristic distributions of the LIDC training, validation and testing sets. Values are shown in mean ± standard deviation.}
\label{LIDC_data_dist}
\begin{tabular}{llll} 
\hline\hline
\textbf{Characteristics} & \multicolumn{1}{c}{\begin{tabular}[c]{@{}c@{}}\textbf{Training set}\\\textbf{(n = 356)}\end{tabular}} & \multicolumn{1}{c}{\begin{tabular}[c]{@{}c@{}}\textbf{Validation set}\\\textbf{~(n = 45)}\end{tabular}} & \multicolumn{1}{c}{\begin{tabular}[c]{@{}c@{}}\textbf{Testing set}\\\textbf{~(n = 492)}\end{tabular}}  \\ 
\hline
Diameter (mm)            & 9.54 ± 4.92                                                                                           & 8.95 ± 5.11                                                                                             & 9.36 ± 4.84                                                                                            \\ 
\hline
Subtlety                 & 3.96 ± 1.10                                                                                           & 3.881 ± 1.074                                                                                           & 3.989 ± 1.096                                                                                          \\ 
\hline
InternalStructure        & 1.05 ± 0.20                                                                                           & 1.00 ± 0.07                                                                                             & 1.020 ± 0.237                                                                                          \\ 
\hline
Calcification            & 5.65 ± 0.96                                                                                           & 5.60 ± 1.032                                                                                            & 5.673 ± 0.934                                                                                          \\ 
\hline
Sphericity               & 3.76 ± 0.99                                                                                           & 3.75 ± 0.988                                                                                            & 3.770 ± 0.989                                                                                          \\ 
\hline
Margin                   & 4.00 ± 1.12                                                                                           & 3.95 ± 1.192                                                                                            & 4.014 ± 1.091                                                                                          \\ 
\hline
Lobulation               & 1.75 ± 1.07                                                                                           & 1.66 ± 1.008                                                                                            & 1.787 ± 1.092                                                                                          \\ 
\hline
Spiculation              & 1.64 ± 1.05                                                                                           & 1.56 ± 0.943                                                                                            & 1.671 ± 1.089                                                                                          \\ 
\hline
Texture                  & 4.58 ± 0.97                                                                                           & 4.50 ± 1.063                                                                                            & 4.614 ± 0.936                                                                                          \\ 
\hline
Malignancy               & 2.99 ± 1.25                                                                                           & 2.99 ± 1.246                                                                                            & 2.991 ± 1.246                                                                                          \\
\hline\hline
\end{tabular}
\end{center}
\textit{Note:} All attributes, except diameter and calcification, are measured on an ordinal scale from 1 to 5, whereas calcification values span from 2 to 6. Spiculation and lobulation indicate the quantity of these formations present within a single nodule. The probabilities of sphericity, calcification, and malignancy in a single nodule are also represented. There is no significant statistical disparity among the characteristics in the three groups.

\end{table}

The overall workflow of preprocessing stage has been demonstrated in Fig. \ref{pre_processing}. The proposed system initiates the automatic segmentation of lung nodules by taking a 2D ROI as input. Our framework first employs the 2D ROI to identify the area to be cropped, with dimensions of $96\times96$, without any resizing. This dimension of patch is selected after statistical analysis of nodule area/appearance on axial axis. Later, the cropped slice patch is transformed into a 2D ROI mask. Lastly, we generate Maximum Intensity Projection (MIP) images of 3 mm in both directions (i.e., forward and backwards). Radiologists commonly use MIP images to identify pulmonary nodules within the lung region. MIP images are generated by overlaying the maximum voxel values from a series of consecutive slices at each coordinate. This process can be mathematically formulated as:

\begin{equation}
\label{mip_eq}
I_{x,y} = \max {\Delta z} D{x,y,\Delta z}
\end{equation}

Here, $I_{x,y}$ and $D_{x,y,\Delta z}$ represent the generated MIP image and the sub-volume or slab of the CT scan, respectively. $\Delta z$ denotes the slab thickness, which in our case is set to 3 mm. To generate the forward and backward MIP images, the range of $\Delta z$ is defined differently and can be formulated as:
\begin{equation}
\begin{aligned}
\Delta z & =\left\{\begin{array}{l}
\left[\phi_c, \phi_c +\phi_{m}\right], \text { For forward MIP image}\\
\left[\phi_c -\phi_{m}, \phi_c\right], \text { For backward MIP image}
\end{array}\right.
\end{aligned}
\end{equation}

In this equation, $\phi_c$ and $\phi_{m}$ represent the depth of the current slice in the scan and the thickness of the MIP to be generated, i.e., 3 mm, respectively. Incorporating MIP images serves to include 3D aspects in both directions, aiding the proposed network in determining bounding boxes for surrounding slices in both directions.

\begin{figure}[b]
\centering
\centerline{\includegraphics[width=.5\textwidth]{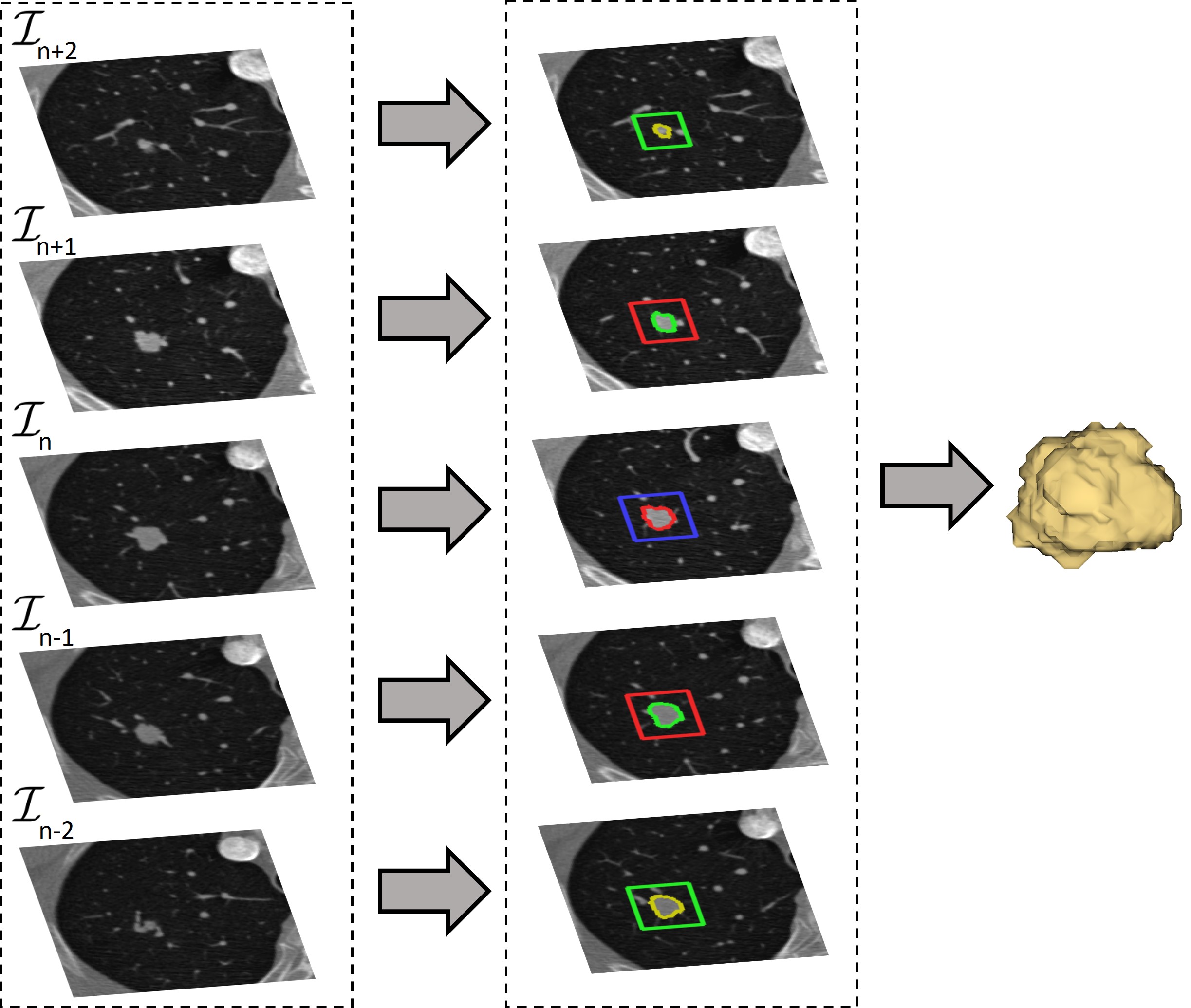}}
\caption{This figure illustrates the successive iterations necessary for complete 3D segmentation of the nodule. The blue region represents the initial provided ROI, while the red and green regions correspond to the outcomes of first and second iterations, respectively. }
\label{iter2}
\end{figure}

\subsection{Training and Inference Strategy}
The proposed model estimates the nodule segmentation mask for the current slice, in addition to the ROI masks for both adjacent slices. The inference process is depicted in the Fig. \ref{iter2}. For generating the ROI masks during training, we leveraged the lung nodule segmentation masks, maintaining the ROI area 30\% larger than the nodule area, which has been proven optimal \cite{usman2020volumetric}. From the 356 training nodules, we treated each slice containing a nodule as a training sample, subsequently generating 3,000 samples with nodules of dimensions $96\times96$. To enhance the model's ability to recognize the absence of a nodule, we incorporated 500 random patches of non-nodular regions with the same dimensions. We adopted a similar strategy for the validation set, producing 800 samples with nodules and 100 samples without nodules. For testing, we provided a single 2D ROI for each nodule to predict the 3D mask, comparing each prediction with the 3D ground truth mask to calculate the overall performance.

We implemented the proposed network using the TensorFlow 2.0 platform. For this experiment, we set the batch size to 15 and optimized the proposed architecture using the Adam optimizer, with an initial learning rate of $10^{-3}$. The network was trained for 200 epochs; despite the large dataset and complex problem, our model parameters were small, resulting in fast convergence of our training curve. We employed binary cross-entropy as the loss function.
\subsection{Evaluation Parameters}
\label{eva}

We utilize six widely-recognized metrics in volume segmentation for evaluation: dice similarity coefficient (DSC, \% ), positive predictive value (PPV, \% ), sensitivity (SEN, \% ), average symmetric surface distance (ASD, $\mathrm{mm}$ ), residual mean square distance (rms, mm), and Hausdorff distance (HFD, mm). The complete definitions are provided in the equations below:
\begin{equation}
\small
\mathrm{DSC}=\frac{2 \mathrm{TP}}{2 \mathrm{TP}+\mathrm{FP}+\mathrm{FN}}
\end{equation}
and
\begin{equation}
\small
\mathrm{PPV}=\frac{\mathrm{TP}}{\mathrm{TP}+\mathrm{FP}}, \quad \mathrm{SEN}=\frac{\mathrm{TP}}{\mathrm{TP}+\mathrm{FN}}
\end{equation}
Here, TP represents the number of accurately segmented lung nodule voxels, FP is the count of falsely segmented lung nodule voxels, and FN stands for the number of falsely segmented background voxels. Additionally,
\begin{align}
\mathrm{ASD} & =\frac{\left(\sum_{z \in S E G} \mathrm{~d}(\mathrm{z}, \mathrm{GT})+\sum_{z \in \mathrm{GT}} \mathrm{d}(\mathrm{z}, \mathrm{SEG})\right)}{|\mathrm{SEG}|+|\mathrm{GT}|} \\
\small\mathrm{rms} & =\sqrt{\frac{\left(\sum_{z \in \mathrm{SEG}} \mathrm{d}(\mathrm{z}, \mathrm{GT})^2+\sum_{\mathrm{z} \in \mathrm{GT}} \mathrm{d}(\mathrm{z}, \mathrm{SEG})^2\right)}{|\mathrm{SEG}|+|\mathrm{GT}|}}
\end{align}
and
\begin{equation}
\small
\mathrm{HFD}=\max [\mathrm{d}(\mathrm{SEG}, \mathrm{GT}), \mathrm{d}(\mathrm{GT}, \mathrm{SEG})]
\end{equation}
In these equations, GT and SEG denote the sets of voxels on the ground truth surface and the automatically segmented lung nodule surface, respectively. The symbol $\mathrm{d}(\mathrm{x}, \mathrm{Y})$ represents the minimum Euclidean distance from voxel $\mathrm{x} \in \mathrm{X}$ to $\mathrm{Y}$.

Higher DSC, PPV, and SEN values signify improved performance, while lower ASD, rms, and HFD values indicate better performance. DSC, PPV, and SEN are commonly employed to gauge the performance of lung nodule segmentation techniques. Furthermore, to thoroughly assess the model's segmentation capability, we also employed ASD, rms, and HFD for detailed analysis.

\section*{Results and Discussion}
\label{RD}
We conducted several experiments on the LIDC-IDRI dataset to assess the effectiveness of our method. The outcomes were examined in terms of overall performance, computational time analysis, robustness, and visual performance. The subsequent subsections discuss our findings in detail.
\begin{table}
\centering
\caption{Average DSC percentages from pairwise comparisons between each of the four radiologists (R1 to R4), ground truth generated with 50\% consensus, and MESAHA-Net.}
\begin{tabular}{llllll} 
\hline\hline
                                                                       & \textbf{R1}   & \textbf{R2}   & \textbf{R3}   & \textbf{R4}   & \textbf{Average}  \\ 
\hline\hline
R1                                                                     & \textit{--}   &    77.16           &       77.86           &           77.44  &        77.48 ± 15.08           \\ 
\hline
R2                                                                     &        77.16       & \textit{--}   &       79.25        &      76.60         &         77.67 ± 14.18          \\ 
\hline
R3                                                                     &        77.86       &       79.25        & \textit{--}   &      77.78         &             78.29 ± 14.60      \\ 
\hline
R4                                                                     &        77.44       &       76.60        &      77.78         & \textit{--}   &         77.27 ± 15.62          \\ 
\hline
\begin{tabular}[c]{@{}l@{}}Consensus (50\%)\\Ground Truth\end{tabular} & 86.47 & 84.63 & 86.68 & 87.48 &  86.31 ± 14.03     \\ 
\hline
MESAHA-Net                                                             & 85.15 &  86.02 & 84.93 & 85.68 & 85.44 ± 17.21     \\
\hline
\end{tabular}
\label{intra_rad_tab}
\end{table}

\begin{figure*}[!ht]
\centering
\centerline{\includegraphics[width=1\textwidth]{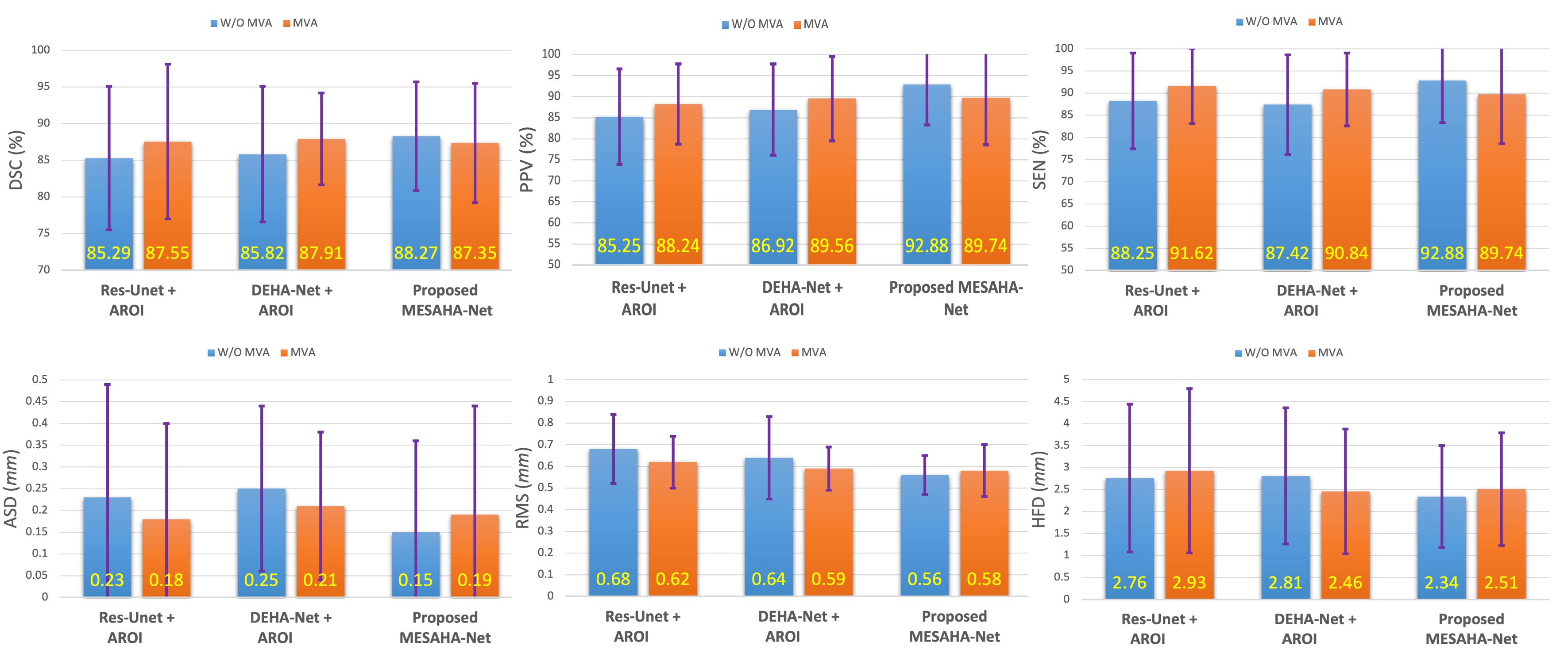}}
\caption{The Dice Similarity Coefficient (DSC), Positive Predictive Value (PPV), Sensitivity (SEN), Average Symmetric Surface Distance (ASD), root mean square (rms), and Hausdorff Fractal Dimension (HFD) values obtained by Res-Unet \cite{usman2020volumetric}, DEHA-Net \cite{usman2022deha}, and our proposed MESAHA-Net, both without and with multi-view analysis (MVA), are presented alongside their respective standard deviations.}
\label{A_result}
\end{figure*}

\subsection*{Overall Performance}
\label{overall_sec}
We analyzed the performance of the proposed framework quantitatively while comparing the results with previously established techniques with respect to DSC, SEN and PPV as these parameters are utilized by most of the studies. Table \ref{comparison_tab} shows the comparison of between our proposed schemes and other state-of-the-art methods. The results demonstrate proposed MESAHA-Net outperforms all the previous studies in term of DSC and SEN. However, DEHA-Net \cite{usman2022deha} obtained slightly better PPV by utilizing the full slice as input with handicraft AROI algorithm.  In comparison with other systems, Res-UNet \cite{usman2020volumetric} and DEHA-Net \cite{usman2022deha} achieved significantly improved performance which can be attributed to the multi-view, i.e., axial, sagittal and coronal, analysis with handicraft AROI algorithm which adds additional computational expense. Similarly, CMSF \cite{zhou2022cascaded} achieved comparative results by employing multiple networks in cascade manner which greatly increases the computational cost and makes it difficult to utilize the solution for the real-time clinical applications. Whereas the proposed scheme only performs axial view analysis with unified architecture which eliminates the need of additional handicraft algorithm to estimate the ROIs for the adjacent slices.  Most importantly, proposed MESAHA-Net eliminates the issues pertaining to image scale normalization by exploiting the adaptive hard attention mechanism.

Although the proposed MESAHA-Net performs patch-wise segmentation along the axial axis, we also explored the possibility of further improving performance by utilizing multi-view analysis (MVA). To this end, we performed segmentation along the sagittal and coronal axes using the proposed MESAHA-Net, and the final segmentation results were produced by employing the max consensus criterion \cite{warfield2004simultaneous}. For the sagittal and coronal views, we utilized fixed ROIs similar to previous methods, such as Res-UNet \cite{usman2020volumetric} and DEHA-Net \cite{usman2022deha}. We compared both versions of our proposed technique, with and without MVA, to previously published studies \cite{usman2020volumetric,usman2022deha} using all evaluation metrics defined in Section \ref{eva}. In particular, the DSC, PPV, SEN, ASD, rms, and HFD values obtained by Res-UNet \cite{usman2020volumetric}, DEHA-Net \cite{usman2022deha}, and our proposed MESAHA-Net without and with MVA are presented in Fig. \ref{A_result}, respectively.

The results indicate that both previous methods perform better with MVA. However, MVA slightly degrades the performance of the proposed MESAHA-Net, which can be attributed to the poor performance of the proposed architecture along the sagittal and coronal axes due to the incorporation of multiple inputs. In particular, the higher slice thickness diminishes the effectiveness of forward and backward MIP images when performing segmentation along sagittal and coronal views. This finding suggests that the proposed architecture does not require MVA, as the incorporation of MIP images compensates for 3D contextual information, which assists in accurate nodule segmentation.

The LIDC/IDRI dataset, annotated by four radiologists, employs a 50\% consensus criterion to determine the ground truth, as illustrated in Table \ref{comparison_tab}. To perform an insightful analysis concerning radiologist agreement, we conducted a pairwise DSC assessment among all radiologists, the consensus ground truth, and the output of MESAHA-Net on the test set. The results are illustrated in Table \ref{intra_rad_tab}. The results indicate that the average DSC between the consensus ground truth and each radiologist is 86.31\%, which is considerably greater than the average inter-radiologist variability of 78.29\%. Since the proposed MESAHA-Net is trained using ground truth, it exhibits consistency when compared to the four individual radiologists. The average DSC between MESAHA-Net and each radiologist remains stable within the range of 84.93\%–85.68\%.

\begin{table}[t]
\centering
\caption{The average $\pm$ standard deviation for the quantitative outcomes of different segmentation techniques, with the best performance highlighted in bold.}
\centering
\scalebox{0.9}{
\begin{tabular}{llll}
\multicolumn{1}{c}{\textbf{Methodology, year~}} & \multicolumn{1}{c}{\textbf{DSC (\%)}} & \multicolumn{1}{c}{\textbf{SEN (\%)}} & \multicolumn{1}{c}{\textbf{PPV (\%)}}  \\ 
\hline
CF-CNN \cite{wang2017central}, 2017             & 82.15$\pm$10.76                       & 92.75$\pm$12.83                       & 75.84$\pm$13.14                        \\ 
\hline
CDP-ResNet + IWS \cite{liu2019cascaded}, 2019                  & 81.58$\pm$11.05                       & 87.30$\pm$14.30                       & 79.71$\pm$13.59                        \\ 
\hline
Res-UNET + AROI \cite{usman2020volumetric}, 2020              & 87.55$\pm$10.58                       & 91.62$\pm$8.47                        & 88.24$\pm$9.52                         \\ 
\hline
DB-ResNet \cite{cao2020dual}, 2020              & 82.74$\pm$10.19                       & 89.35$\pm$11.79                       & 79.64$\pm$13.34                        \\ 
\hline
DB UNET + LLIE \cite{wu2021coarse}, 2021        & 81.97$\pm$10.16                       & 84.57$\pm$14.79                       & 79.33$\pm$14.08                        \\ 
\hline
MTGAN \cite{chen2021mtgan}, 2021        & 85.24$\pm$9.01                       & \textit{--}                      & \textit{--}                        \\ 
\hline
M-SegSEUNet-CRF \cite{zhang2022multi}, 2022     & 80.50$\pm$7.6                          & 79.80$\pm$14.30                        & 84.6$\pm$10.9                          \\ 
\hline
CSE-GAN \cite{tyagi2022cse}, 2022               & 80.74$\pm$ \textit{--}                 & 85.46$\pm$ \textit{--}                 & 80.56$\pm$ \textit{--}                  \\ 
\hline
CMSF \cite{zhou2022cascaded}, 2022                 & 86.75$\pm$ 9.25                 & 89.07$\pm$ 8.31                 & 83.26$\pm$ 10.21                  \\ 
\hline
FMGA \cite{chen2021multi}, 2022                 & 81.32$\pm$ \textit{--}                 & 92.33$\pm$ \textit{--}                 & 74.78$\pm$ \textit{--}                  \\ 
\hline
DEHA-Net + AROI \cite{usman2022deha}, 2023      & 87.91$\pm$6.27                        & 90.84$\pm$8.22                        & \textbf{89.56$\pm$10.07}                        \\ 
\hline
Our Method~                                     & \textbf{88.27$\pm$7.42}~                       & \textbf{92.88$\pm$9.54}                        & 86.95$\pm$11.29                        \\
\hline
\end{tabular}
}
\label{comparison_tab}
\end{table}

\begin{table}[!ht]
\centering
\caption{DSCs on different nodule groups of the LIDC testing set.}
\begin{tabular}{lllllll} 
\hline\hline
                               & \multicolumn{6}{l}{\textbf{Characteristic scores}}         \\ 
\cline{2-7}
\textbf{Characteristics}       & 1       & 2       & 3       & 4       & 5       & 6        \\ 
\hline
\multirow{2}{*}{Calcification} & 87.38   & $-$   & 88.15   & 81.24   & 80.44   & 88.16    \\ 
                               & {[}1]   & $-$   & {[}49]  & {[}4]   & {[}1]   & {[}438]  \\ 
\hline
Internal structure             & 88.08   & 86.49   & $-$   & 89.61   & $-$   & $-$    \\
                               & [489] & {[}1]   & $-$   & {[}3]   & $-$   & $-$    \\ 
\hline
Lobulation                     & 91.23   & 80.84   & 92.12   & 80.95   & 91.44   & $-$    \\
                               & {[}272] & {[}122] & {[}49]  & {[}32]  & {[}18]  & $-$    \\ 
\hline
Malignancy                     & 83.07   & 94.13   & 86.78   & 83.4    & 92.9    & $-$    \\
                               & {[}70]  & {[}100] & {[}162] & {[}86]  & {[}75]  & $-$    \\ 
\hline
Margin                         & 89.42   & 81.86   & 85.01   & 93.56   & 85.66   & $-$    \\
                               & {[}18]  & {[}39]  & {[}65]  & {[}167] & {[}204] & $-$    \\ 
\hline
Sphericity                     & 87.71   & 93.45   & 81.85   & 90      & 91.26   & $-$    \\
                               & {[}4]   & {[}44]  & {[}155] & {[}148] & [142]  & $-$    \\ 
\hline
Speculation                    & 86.72   & 94.62   & 85.29   & 78.38   & 92.6    & $-$    \\
                               & {[}312] & {[}96]  & {[}43]  & {[}19]  & {[}23]  & $-$    \\ 
\hline
Subtlety                       & 85.03   & 92.36   & 93.15   & 89.71   & 84.2    & $-$    \\
                               & {[}14]  & {[}39]  & {[}98]  & {[}129] & {[}213] & $-$    \\ 
\hline
Texture                        & 93.54   & 94.13   & 93.6    & 83.91   & 87.93   & $-$    \\
                               & {[}20]  & {[}5]   & {[}23]  & {[}49]  & {[}396] & $-$    \\
\hline\hline
\end{tabular}
\label{robustness_tab}
\end{table}

\begin{figure}[b]
\centering
\centerline{\includegraphics[width=.4\textwidth]{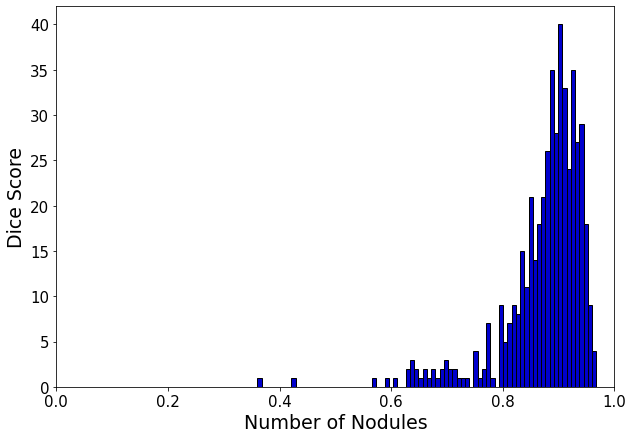}}
\caption{Distribution of DSC for the LIDC-IDRI test set.}
\label{hist_result}
\end{figure}

\subsection*{Robustness Analysis}
\label{robust_sec}
The LIDC-IDRI dataset also contains annotations about several nodule features, including subtlety, internal structure, calcification, sphericity, margin, lobulation, spiculation, texture, and malignancy. These characteristics reflect the varying difficulty levels in defining nodule boundaries. To evaluate the robustness of our method, we separated the test data according to the different levels of each feature and extracted the results for each level, as displayed in Table \ref{robustness_tab}. The findings imply that the dice scores attained for each group are similar, indicating the robustness of our approach.

Additionally, we depicted the distribution of dice scores achieved on the LIDC-IDRI dataset's test set, as shown in Fig. \ref{hist_result}. The outcomes reveal that most of the test instances have scores above 80





\begin{figure}[t]
\centering
\centerline{\includegraphics[width=.5\textwidth]{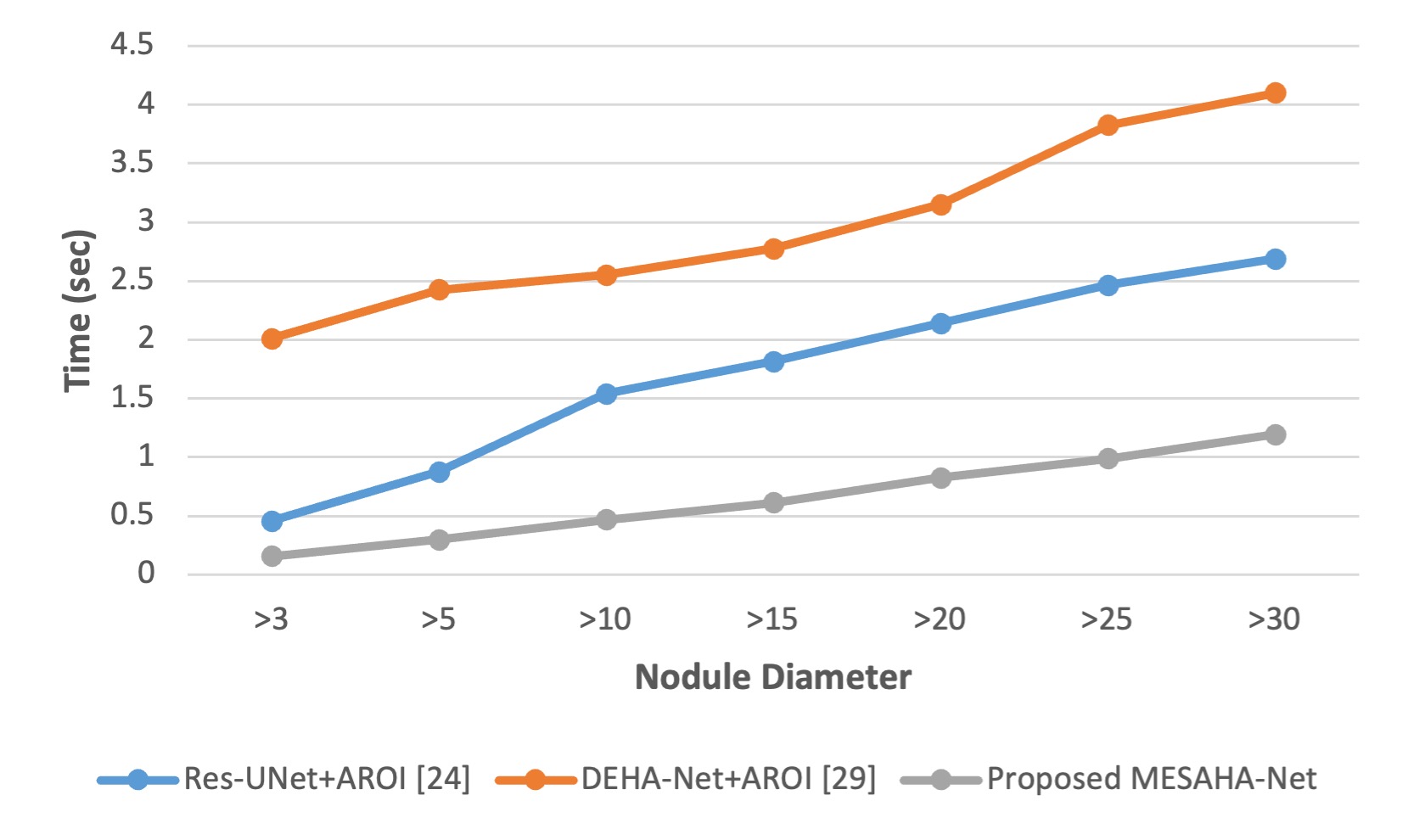}}
\caption{Relationship between nodule size and the computation time taken by three lung nodule segmentation schemes.}
\label{time_graph}
\end{figure}

\begin{table}
\centering
\caption{Comparison of different networks }
\begin{tabular}{lcc} 
\hline\hline
\textbf{Network}         & \begin{tabular}[c]{@{}c@{}}\textbf{Number of }\\\textbf{parameters}\end{tabular} & \begin{tabular}[c]{@{}c@{}}\textbf{Inference time }\\\textbf{per slice (ms)}\end{tabular}  \\ 
\hline\hline
VNet (3D) \cite{milletari2016v}             & 9,448,849                                                                        & 127.95                                                                                     \\ 
\hline
UNet (3D) \cite{cciccek20163d}            & 8,822,065                                                                        & 119.46                                                                                     \\ 
\hline
ResUNet (2D) \cite{usman2020volumetric}          & 6,272,916                                                                        & 96.61                                                                                      \\ 
\hline
DEHA-Net (2D) \cite{usman2022deha}         & 536,447                                                                          & 72.84                                                                                      \\ 
\hline
Proposed MESAHA-Net (2D) & 436,707                                                                          & 48.34                                                                                      \\
\hline
\end{tabular}
\label{comparison_params}
\end{table}

\subsection*{Computational Time Analysis}
\label{timesec}
In real-time clinical settings, the time required for lung nodule segmentation is of paramount importance. Consequently, the computation time of automatic segmentation solutions plays a critical role in evaluating the efficiency of lung nodule segmentation systems. In Table \ref{comparison_tab}, we compared previously published studies, among which Res-UNet with A-ROI\cite{usman2020volumetric} and DEHA-Net \cite{usman2022deha} achieved comparable performance. Both techniques employ a similar patch-wise investigation strategy, taking a single ROI as input to generate 3D segmentation of lung nodules. Subsequently, in this section, we present a comparison of the computational time of the proposed MESAHA-Net with Res-UNet with A-ROI\cite{usman2020volumetric} and DEHA-Net \cite{usman2022deha}. To ensure a fair analysis, we implemented all three schemes on the same hardware, i.e., an NVIDIA TITAN RTX GPU with a 32-core Intel Asus CPU. To perform the computational time analysis, we first divided the entire test set into seven groups based on lung nodule diameter (i.e., 3--5, 5--10, 10--15, 15--20, 20--25, 25--30, and greater than 30 mm).

We utilized the same ROI for the initialization of segmentation by each technique and recorded the total computational time taken by each technique to produce 3D segmentation of each lung nodule. Fig. \ref{time_graph} demonstrates the average times recorded for each group of nodules with all three techniques. Firstly, all three techniques follow an increasing trend with respect to the increase in diameter, which is due to the patch-wise (2D) nature of these systems. Larger nodules usually cover more slices, and for performing slice-by-slice segmentation of nodules, more inferences are required. However, the increase in computational time is significantly sharp for previously published studies, whereas the proposed MESAHA-Net demonstrates an incremental change in computational time and performs 3D segmentation of lung nodules more efficiently.

We also compared the number of parameters for each network in Table \ref{comparison_params}, where we included two 3D models as baselines, i.e., VNet \cite{milletari2016v} and UNet \cite{cciccek20163d}. We utilized all the techniques in the table to make inferences on the test set and calculate the inference time per slice. The results indicate that the proposed MESAHA-Net achieved the shortest inference time, making it the most efficient technique for lung nodule segmentation. This optimal performance can be attributed to the following factors: (1) The proposed MESAHA-Net is a lightweight model with fewer parameters compared to Res-UNet\cite{usman2020volumetric} and DEHA-Net\cite{usman2022deha}. (2) The proposed technique only performs axial view analysis, whereas the other two methods necessitate multi-view analysis to achieve similar performance. (3) The proposed solution employs a unified single model, i.e., MESAHA-Net, eliminating the need for additional handcrafted algorithms, such as A-ROI, to estimate the ROIs for preceding and subsequent slices. (4) Unlike DEHA-Net \cite{usman2022deha}, which takes a full slice as input to circumvent rescaling-induced errors, the proposed MESAHA-Net only takes a patch of a slice, i.e., $96\times96$, as input.

\subsection*{Qualitative Analysis}
\begin{figure*}[!ht]
\centering
\centerline{\includegraphics[width=0.9\textwidth]{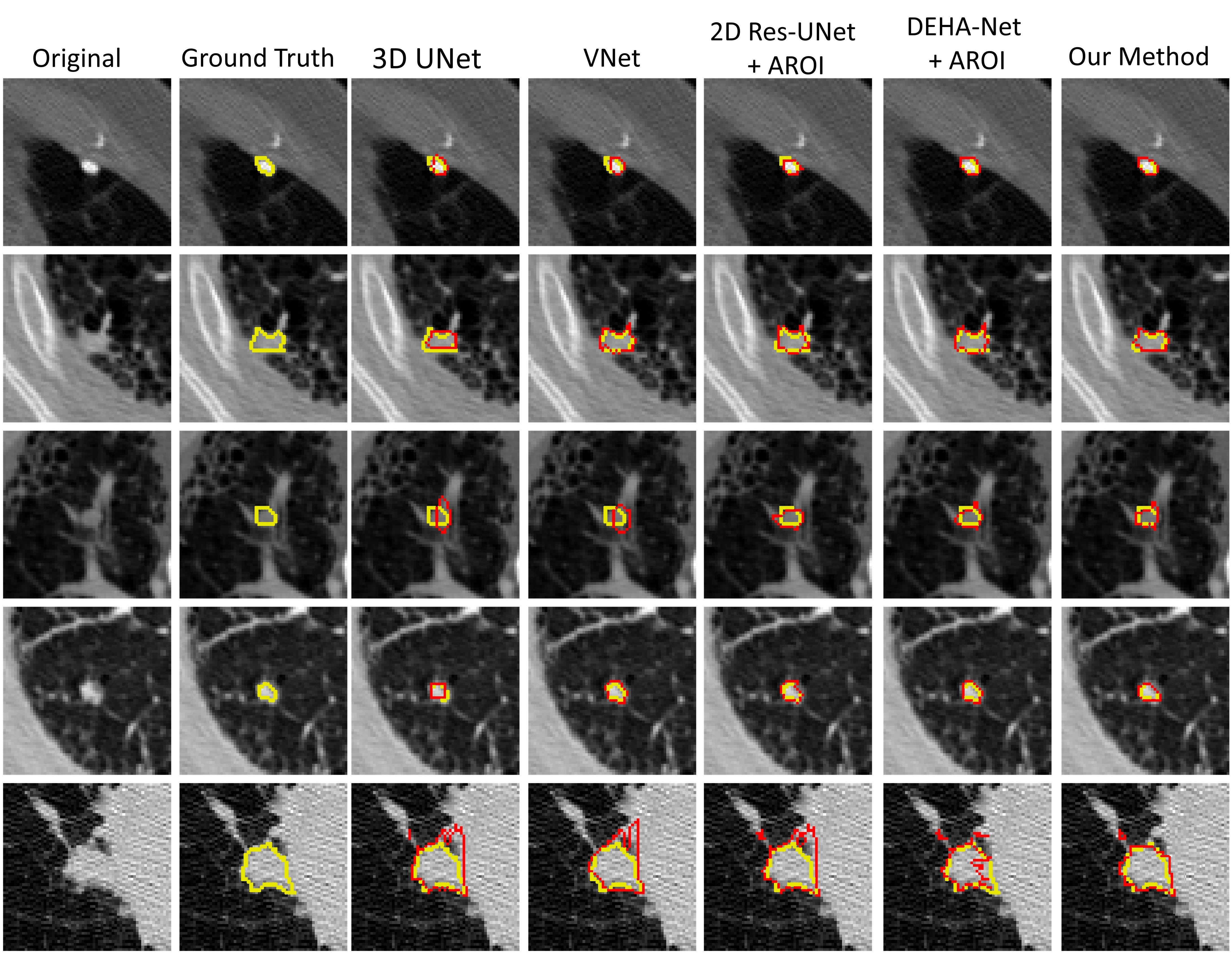}}
\caption{A visual comparison of the results obtained from 3D UNet\cite{cciccek20163d}, VNet\cite{milletari2016v}, 2D ResUNet\cite{usman2020volumetric}, 2D DEHA-Net \cite{usman2022deha} with the AROI algorithm, and our proposed MESAHA-Net is presented. Each row depicts the central slice of different nodule from test set.}
\label{v_results_2D}
\end{figure*}

In this section, we present a qualitative analysis of the results obtained from various methods for different types and sizes of lung nodules. Fig. \ref{v_results_2D} illustrates the 2D segmentation results from five distinct techniques, including the proposed MESAHA-Net, on five diverse types and sizes of lung nodules, i.e., L1--L5. Overall, it can be observed that 2D patch-wise approaches yield better results than 3D network-based methods, primarily due to the significant variations in lung nodule sizes. The results also demonstrate that the proposed technique surpasses other methods while maintaining consistent performance across different types and sizes of nodules. This suggests that the proposed MESAHA-Net effectively improves lung nodule segmentation by leveraging self-adaptive hard attention to mitigate rescaling-induced errors.

\begin{figure}[!ht]
\centering
\centerline{\includegraphics[width=0.5\textwidth]{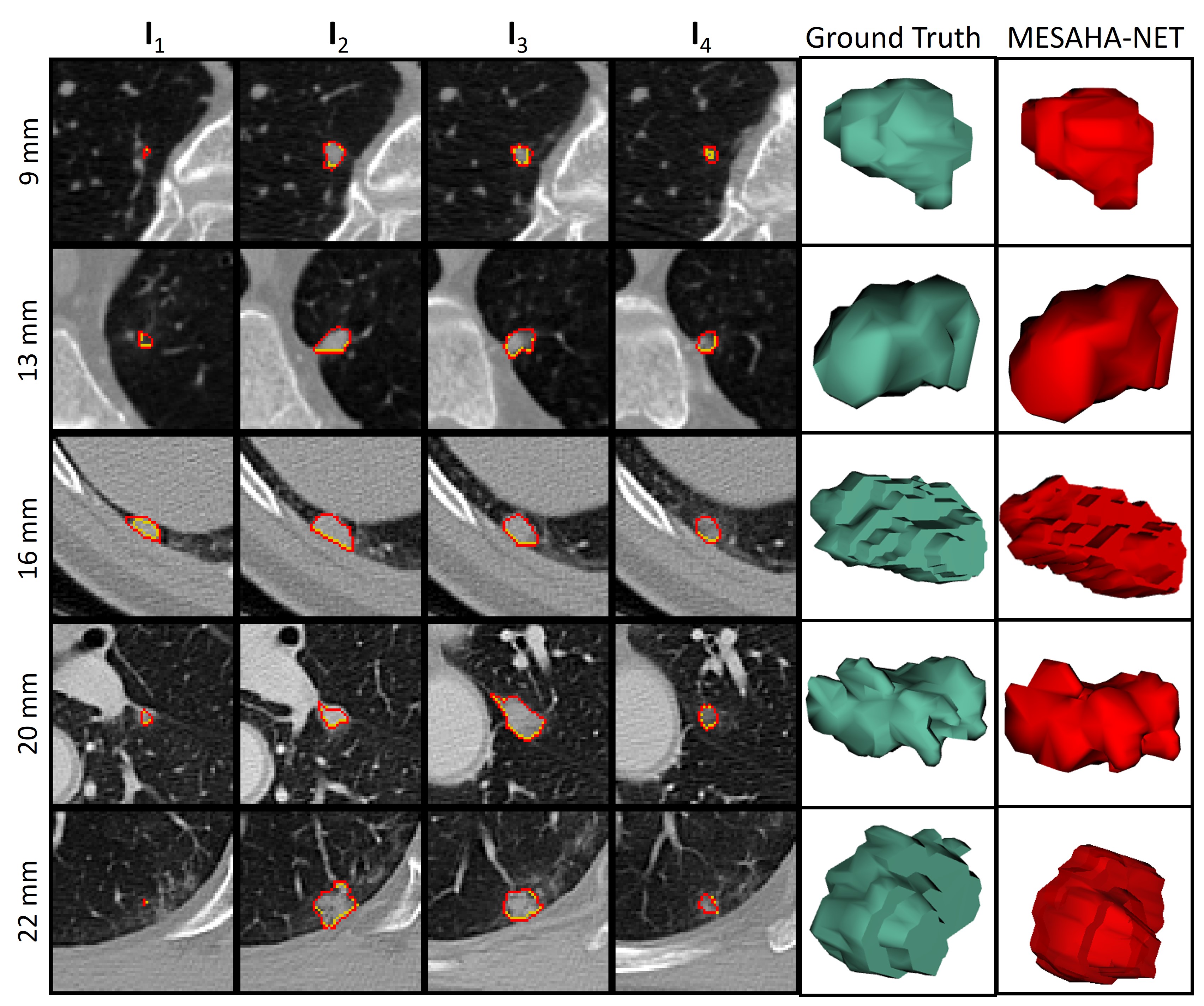}}
\caption{The segmentation results of the proposed MESAHA-Net on various nodules of differing sizes are displayed. Each row represents slices from a single nodule within the test set.}
\label{v_result}
\end{figure}

To further evaluate the performance of the proposed scheme in detecting the presence of nodules in each slice, we conducted a 3D rendering of the final 3D segmentation for various nodules. Fig. \ref{v_result} presents the results obtained for five lung nodules with different diameters, showcasing multiple segmented slices and the 3D rendered outcomes of the proposed MESAHA-Net alongside the ground truth. These results demonstrate that the proposed scheme effectively identifies the extent of lung nodules in preceding and subsequent slices, enabling accurate 3D segmentation of lung nodules.

\section*{Conclusion}
\label{co}
In this study, we presented a novel multi-encoder-based self-adaptive hard attention network (MESAHA-Net) that achieves accurate and efficient segmentation of pulmonary nodules. Our proposed framework perform patch-wise segmentation of lung nodules iteratively by focusing on the nodule region in each slice to generate 3D volumetric segmentation of lung nodules. The MESAHA-Net utilizes three distinct inputs: raw slice patch, bidirectional maximum intensity projection (MIP) images, and hard attention in the form of regions of interest (ROI) masks. This incorporation enables the framework to harness both 2D and 3D spatial and contextual information, essential for accurately segmenting lung nodules and estimating ROIs for neighboring slices. Additionally, we have introduced an adaptive hard-attention mechanism that enforces the network to focus on the nodule region while mitigating rescaling-induced errors. We rigorously evaluated our proposed method on the LIDC-IDRI dataset, the most comprehensive benchmark dataset available. The findings indicate that our framework surpasses previously published lung nodule segmentation approaches in terms of performance and computational complexity, rendering it suitable for real-time clinical applications.

\end{document}